\documentclass[11pt,a4paper]{article}

\pdfoutput=1

\usepackage{jheppub}
\usepackage{bbold}
\usepackage{amsfonts}
\usepackage{cleveref}
\usepackage{color}

\makeatletter
\def\@fpheader{Prepared for submission to JCAP}
\makeatother

\newcommand{\sdgb}{\sqrt{-g_\eff}}
\newcommand{\mn}{{\mu\nu}}
\newcommand{\ab}{{ab}}
\newcommand{\calL}{\mathcal{L}_m}
\newcommand{\eff}{\mathrm{eff}}

\makeatletter
\DeclareRobustCommand{\rcite}[1]{%
  \rcite@aux#1,\@nil{#1}%
}
\def\rcite@aux#1,#2\@nil#3{%
  \if\relax#2\relax
    Ref.~\cite{#3}%
  \else
    Refs.~\cite{#3}%
  \fi
}
\makeatother

\title{Cosmic expansion histories in massive bigravity with symmetric matter coupling}

\author[a]{Jonas Enander,}
\author[b]{Adam R. Solomon,}
\author[c]{Yashar Akrami,}
\author[a]{and Edvard M\"ortsell}
\affiliation[a]{Oskar Klein Center, Stockholm University,\\Albanova University Center\\ 106 91 Stockholm, Sweden}
\affiliation[b]{DAMTP, Centre for Mathematical Sciences, University of Cambridge\\ Wilberforce Rd., Cambridge CB3 0WA, UK}
\affiliation[c]{Institute of Theoretical Astrophysics, University of Oslo\\ P.O. Box 1029 Blindern, N-0315 Oslo, Norway}
\emailAdd{enander@fysik.su.se}
\emailAdd{a.r.solomon@damtp.cam.ac.uk}
\emailAdd{yashar.akrami@astro.uio.no}
\emailAdd{edvard@fysik.su.se}
\abstract{We study the cosmic expansion history of massive bigravity with a viable matter coupling which treats both metrics on equal footing. We derive the Friedmann equation for the effective metric through which matter couples to the two metrics, and study its solutions. For certain parameter choices, the background cosmology is identical to that of $\Lambda$CDM. More general parameters yield dynamical dark energy, which can still be in agreement with observations of the expansion history. We study specific parameter choices of interest, including minimal models, maximally-symmetric models, and a candidate partially-massless theory.}


\begin{document}

\maketitle
 
\section{Introduction}

Massive gravity and its bimetric extension, massive bigravity, provide a ghost-free generalization of general relativity in which the graviton is endowed with a nonvanishing mass \cite{deRham:2010ik,deRham:2010kj,deRham:2011rn,deRham:2011qq,
Hassan:2011vm,Hassan:2011hr,Hassan:2011tf,Hassan:2011zd,deRham:2014zqa}. These theories are carefully constructed to render nondynamical the Boulware-Deser (BD) ghost mode which plagues most nonlinear theories of a massive graviton \cite{Boulware:1973my}. Massive gravity necessarily introduces a second metric, often called the reference metric. In single-metric massive gravity the reference metric is fixed (which is usually referred to as the de Rham-Gabadadze-Tolley theory, or dRGT), while in the bimetric extension the second metric is dynamical (also known as the Hassan-Rosen theory).

Depending on how matter couples to the two metrics, there might be ambiguities when it comes to light propagation and physical observables. These issues have been demonstrated, for example, in \cite{Akrami:2013ffa,Akrami:2014lja} for the case where matter couples to both metrics minimally. Such ambiguities, if present, are independent of whether the theory is healthy or contains ghosts. However, most worryingly, many of such doubly-coupled theories, including the one studied in \cite{Akrami:2013ffa,Akrami:2014lja}, reintroduce the BD ghost \cite{Hassan:2012wr,Yamashita:2014fga,deRham:2014naa}.

In a recent paper \cite{deRham:2014naa}, a matter coupling to both metrics was proposed which was shown to be ghost-free around cosmological backgrounds \cite{deRham:2014fha} (the status of the ghost in this specific coupling has also been investigated in \rcite{Hassan:2014gta,deRham:2014fha,Noller:2014ioa,Soloviev:2014eea}). This is achieved by coupling matter minimally to an effective metric constructed out of the two metrics appearing in the gravitational sector of the theory, regardless of whether the second metric is dynamical. This would alleviate the problem of constructing physical observables, as matter would move on geodesics of the effective metric. The proposal of \rcite{deRham:2014naa} has been derived using complementary methods and extended to a multi-metric framework in \rcite{Noller:2014sta}. The cosmology of this new coupling was first investigated in the dRGT context in \rcite{deRham:2014naa}, and is further studied in \rcite{Solomon:2014iwa,Gumrukcuoglu:2014xba}.

Phenomenological studies of singly-coupled bigravity have shown that the cosmological expansion and spherically-symmetric solutions give viable alternatives to general relativity \cite{Volkov:2011an,Comelli:2011zm,vonStrauss:2011mq,Berg:2012kn,Akrami:2012vf,Akrami:2013pna,Konnig:2013gxa,Enander:2013kza,Solomon:2014dua}, but that instabilities arise at the perturbative level (as first shown in \rcite{Comelli:2012db}, with \cite{Konnig:2014xva,Comelli:2014bqa,DeFelice:2014nja,Konnig:2014dna,Lagos:2014lca} studying the problem in further detail and showing specific submodels and parameter choices that can avoid the instabilities). The singly-coupled theory also spoils the metric interchange symmetry present in vacuum; the kinetic and mass terms treat the metrics on equal footing, but this is broken when one couples matter to only one metric. It is therefore important to investigate other types of matter coupling that retain the metric interchange symmetry. The notion of double coupling---in which the same matter is coupled to both metrics, or each metric couples to a different type of matter---has been studied in \rcite{Akrami:2013ffa,Akrami:2014lja,Comelli:2012db,Tamanini:2013xia,DeFelice:2014nja,Aoki:2014cla,Heisenberg:2014rka,Afshar:2014dta}.

In this paper, we study the background cosmology of massive bigravity when matter couples to the effective metric proposed in \rcite{deRham:2014naa}. We show that the background expansion can asymptotically approach $\Lambda$CDM at both early and late times, and for certain parameter values is identical to $\Lambda$CDM always. At the background level, this type of coupling is therefore consistent with observations. In a future study, we will investigate whether this holds true for cosmological perturbations.

In \cref{sec:Action} we present the effective metric and the symmetries that are present in the action. In \cref{sec:flrw} we derive the cosmological equations of motion and discuss their main features. A parameter scan of the minimal models, where only one of the interaction terms is nonvanishing, is performed in \cref{sec:minimod}. In \cref{sec:specialparams} we discuss some special parameter choices. We conclude in \cref{sec:Conclusions}.

\section{Effective metric and coupling}
\label{sec:Action}

The doubly-coupled bigravity action is given by\footnote{Our convention for the Minkowski metric is $(-,+,+,+)$.}
\begin{align}
S & = -\frac{M_{g}^{2}}{2}\int d^{4}x\sqrt{-\det g}R\left(g\right)-\frac{M_{f}^{2}}{2}\int d^{4}x\sqrt{-\det f}R\left(f\right) \nonumber \\
 & \hphantom{{}=}+m^{4}\int d^{4}x\sqrt{-\det g}\sum_{n=0}^{4}\beta_{n}e_{n}\left(\sqrt{g^{-1}f}\right) \nonumber \\
 & \hphantom{{}=} +\int d^{4}x\sqrt{-\det g_\eff}\mathcal{L}_{m}\left(g_\eff,\Phi\right). \label{eq:actionunscaled}
\end{align}
Here $g$ and $f$ are rank-2 tensor fields with Lorentzian signature, $R\left(g\right)$ and $R\left(f\right)$ are their respective Ricci scalars, and $e_{n}$ are the elementary symmetric polynomials presented in \rcite{Hassan:2011vm}. The effective metric, first introduced in \rcite{deRham:2014naa}, is defined by\footnote{In \rcite{deRham:2014naa} the effective metric is given in an explicitly symmetric form, but this is not needed since $g_{\mu\alpha}X^{\alpha}_{\nu} = g_{\nu\alpha}X^{\alpha}_{\mu}$, as first shown in \rcite{Hassan:2012wr}; see also \cref{app:sym}.}
\begin{equation}
\label{eq:effectiveg}
g^\eff_{\mu\nu} = \alpha^2 g_{\mu\nu} + 2\alpha\beta g_{\mu\alpha}X^{\alpha}_{\nu}+\beta^2 f_{\mu\nu},\qquad X^{\mu}_{\nu} = (\sqrt{g^{-1}f})^\mu_\nu.
\end{equation}
As shown in \cref{app:sym}, the effective metric is symmetric under the interchange $g_\mn\leftrightarrow f_\mn$ and $\alpha\leftrightarrow \beta$. This makes the entire action symmetric under the transformations
\begin{equation}
g_\mn\leftrightarrow f_\mn,\qquad M_g \leftrightarrow M_f,\qquad \beta_{n}\rightarrow \beta_{4-n},\qquad\alpha\leftrightarrow\beta.
\end{equation}
There is thus a duality between the two metrics present in the action which is spoiled when matter couples to only one of the metrics (taken by setting either $\alpha=0$ or $\beta=0$). Note that this duality could be ruined above the energy scale where the ghost enters.

The action (\ref{eq:actionunscaled}) contains two Planck masses ($M_g$ and $M_f$), five interaction parameters ($\beta_n$, of which $\beta_0$ and $\beta_4$ are the cosmological constants for $g$ and $f$, respectively), and two parameters describing how matter couples to each metric ($\alpha$ and $\beta$). The Planck masses and the coupling parameters $\alpha$ and $\beta$ only enter observable quantities through their ratios. Moreover, one of those ratios is redundant: as described in appendix \ref{app:sym}, the action can be freely rescaled so that either $M_f/M_g$ or $\beta/\alpha$ is set to unity. Therefore the physically-relevant parameters are $\beta_n$ and either $M_f/M_g$ or $\beta/\alpha$. In this paper we will rescale the Planck masses so that there is one effective gravitational coupling strength, $M_\eff$. We will also keep $\alpha$ and $\beta$ explicit to make the $\alpha \leftrightarrow \beta$ symmetry manifest, but the reader should bear in mind that only their ratio matters physically. All observational constraints will be given solely in terms of $\beta/\alpha$.

The Einstein equations have been derived in \rcite{Schmidt-May:2014xla} and can be written in the form
\begin{align}
(X^{-1})^{(\mu}{}_\alpha G^{\nu)\alpha}_g + m^2\displaystyle\sum_{n=0}^{3}(-1)^n\beta_ng^{\alpha\beta}(X^{-1})^{(\mu}{}_\alpha Y^{\nu)}_{(n)\beta} = \frac{\alpha}{M_\eff^2}\sqrt{\frac{\det g_\eff}{\det g}}\left(\alpha (X^{-1})^{(\mu}{}_\alpha T^{\nu)\alpha}+\beta T^{\mu\nu}\right),\label{eq:eomg}
\end{align}
\begin{align}
X^{(\mu}{}_\alpha G^{\nu)\alpha}_f + m^2\displaystyle\sum_{n=0}^{3}(-1)^n\beta_{4-n}f^{\alpha\beta}X^{(\mu}{}_\alpha \hat{Y}^{\nu)}_{(n)\beta} = \frac{\beta}{M_\eff^2}\sqrt{\frac{\det g_\eff}{\det f}}\left(\alpha T^{\mu\nu}+\beta X^{(\mu}{}_\alpha T^{\nu)\alpha}\right).\label{eq:eomf}
\end{align}
The matrices $Y$ and $\hat{Y}$ depend on $\sqrt{g^{-1}f}$ and $\sqrt{f^{-1}g}$, respectively, and are given in \rcite{Hassan:2011vm}. $G_g^\mn$ and $G_f^\mn$ have their indices raised with $g_\mn$ and $f_\mn$, respectively. The stress-energy tensor $T^\mn$ is defined with respect to the effective metric $g_\eff$ as
\begin{equation}
 \delta\left[\sdgb\calL\left(g_\eff,\Phi\right)\right] = \frac{1}{2}\sdgb T^\mn\delta g^\eff_\mn,
\end{equation}
and obeys the usual conservation equation
\begin{equation}
\nabla^\eff_\mu T^\mn = 0.
\end{equation}

\section{Friedmann equations and their solutions}
\label{sec:flrw}

To describe homogeneous and isotropic cosmologies, we specialize to the  Friedmann-Lema\^itre-Robertson-Walker (FLRW) ans{\"a}tze for both $g_\mn$ and $f_\mn$,
\begin{align}
ds_{g}^{2}&=-N_{g}^{2}dt^{2}+a_{g}^{2}d\vec{x}^{2}, \\
ds_{f}^{2}&=-N_{f}^{2}dt^{2}+a_{f}^{2}d\vec{x}^{2},
\end{align}
where $N_{g,f}$ and $a_{g,f}$ are the lapses and scale factors, respectively, of the two metrics. As in general relativity, we can freely rescale the time coordinate to fix either $N_g$ or $N_f$; however, their ratio is gauge-invariant and will remain unchanged. The effective metric becomes
\begin{equation}
ds_\eff^{2}=-N^{2}dt^{2}+a^{2}d\vec{x}^{2},
\end{equation}
where the effective lapse and scale factor are related to those of the $g$ and $f$ metrics by
\begin{align}
N &= \alpha N_{g} + \beta N_{f}, \\
a &= \alpha a_{g} + \beta a_{f}.
\end{align}
The equations of motion can be derived either directly from \cref{eq:eomg,eq:eomf}, or by plugging the FLRW ans{\"a}tze into the action and varying with respect to the scale factors and lapses. Both approaches yield the same result. The second approach is described in appendix \ref{app:eom}. Here we just state the relevant equations. Defining
\begin{align}
\label{eq:B0}
B_{0}(r)&\equiv\beta_{0}+3\beta_{1}r+3\beta_{2}r^2+\beta_{3}r^3, \\
\label{eq:B1}
B_{1}(r)&\equiv\beta_{1}r^{-3}+3\beta_{2}r^{-2}+3\beta_{3}r^{-1}+\beta_{4},
\end{align}
where
\begin{equation}
r \equiv \frac {a_{f}}{a_{g}},
\end{equation}
the Friedmann equations for the $g$ and $f$ metrics are
\begin{align}
\label{eq:friedeqg}
3H_{g}^2 &= \frac{\alpha\rho}{M_\eff^2}\frac{a^3}{a_{g}^3} + m^2B_0, \\
\label{eq:friedeqf}
3H_{f}^2 &= \frac{\beta\rho}{M_\eff^2}\frac{a^3}{a_{f}^3} + m^2B_1.
\end{align}
Here the energy density $\rho$ is a function of the effective scale factor $a$, and we have defined the $g$- and $f$-metric Hubble rates as
\begin{equation}
H_{g} \equiv \frac{\dot a_{g}}{N_{g}a_{g}}, \qquad H_{f} \equiv \frac{\dot a_{f}}{N_{f}a_{f}}.
\end{equation}
Notice that the two Friedmann equations for $H_g$ and $H_f$ map into one another under the interchange $\beta_n \rightarrow \beta_{4-n}$, $\alpha \leftrightarrow \beta$, and $g_\mn \leftrightarrow f_\mn$ (which sends $H_g \leftrightarrow H_f$, $r\rightarrow r^{-1}$, and $B_0\leftrightarrow B_1$), as expected from the properties of the action described in appendix \ref{app:sym}.

The stress-energy tensor is conserved with respect to the effective metric, so we immediately have
\begin{equation}
\dot{\rho}+3\frac{\dot{a}}{a}\left(\rho+p\right)=0,
\end{equation}
where the density, $\rho$, and pressure, $p$, are defined in the usual way from the stress-energy tensor. By taking the divergence of either Einstein equation with respect to the associated metric (e.g., taking the $g$-metric divergence of \cref{eq:eomg}) and using the Bianchi identity and stress-energy conservation, we obtain the ``Bianchi constraint,"
\begin{equation}
\label{eq:bianchi}
\left[m^{2}\left(\beta_1 a_g^2+2\beta_2 a_g a_f + \beta_3 a_f^2\right)-\frac{\alpha\beta a^{2}p}{M_\eff^2}\right]\left(N_{f}\dot{a}_{g}-N_{g}\dot{a}_{f}\right)=0.
\end{equation}
In complete analogy with the singly-coupled case (obtained by setting $\alpha$ or $\beta$ to zero), \cref{eq:bianchi} gives rise to two possible branches of solutions, one algebraic and one dynamical \cite{Comelli:2011zm,vonStrauss:2011mq,Volkov:2011an}.\footnote{In the singly-coupled theory, \cref{eq:bianchi} would be a constraint equation arising from the Bianchi identity and stress-energy conservation. When using the effective coupling, the stress-energy conservation holds with respect to the effective metric, rather than $g_\mn$ or $f_\mn$. This gives rise to the pressure-dependent term in the left bracket. Due to this term, both branches---obtained by setting either bracket to zero---can be regarded as dynamical. We choose to adopt the terminology from the singly-coupled case here, however.}  In the singly-coupled case, setting the first bracket to zero gives an algebraic constraint on $r$ that can be shown to give solutions that are indistinguishable from general relativity at all scales \cite{vonStrauss:2011mq}. 

In the remainder of this paper we will restrict our study to solutions where the second bracket in \cref{eq:bianchi} vanishes, which is the approach usually taken in singly-coupled bigravity models \cite[e.g.,][]{vonStrauss:2011mq,Berg:2012kn,Akrami:2012vf,Akrami:2013pna,Solomon:2014dua,Konnig:2013gxa,Konnig:2014xva}. The phenomenology of the algebraic branch, which is richer compared to the singly-coupled theory, is discussed in appendix \ref{sec:firstbranch}. In the dynamical branch, we get a constraint on the ratio between $N_f$ and $N_g$:
\begin{equation}
\label{eq:NfNg}
\frac{N_{f}}{N_{g}}=\frac{\dot{a}_{f}}{\dot{a}_{g}}=\frac{da_{f}}{da_{g}}.
\end{equation}
This implies the simple relation $H_{f}r=H_{g}$. Furthermore, the physical Hubble rate $H$, defined as
\begin{equation}
\label{eq:H}
H \equiv \frac{\dot a}{Na},
\end{equation}
becomes
\begin{equation}
H = \frac{H_g}{\alpha+\beta r} = \frac{r H_f }{\alpha+\beta r}.
\end{equation}
Combining the two Friedmann equations, we obtain the evolution equations for $H$ and $r$:
\begin{align}
H^{2}&=\frac{\rho}{6M_{\eff}^{2}}\left(\alpha+\beta r\right)\left(\alpha+\beta r^{-1}\right)+\frac{m^{2}\left(B_{0}+r^{2}B_{1}\right)}{6\left(\alpha+\beta r\right)^{2}}, \label{eq:Heff}\\
0 &= \frac{\rho}{M_\eff^{2}}\left(\alpha+\beta r\right)^{3}\left(\alpha -\beta r^{-1}\right)+m^{2}\left(B_0-r^2B_1\right). \label{eq:quartic}
\end{align}
\Cref{eq:Heff,eq:quartic} determine the expansion history completely and are invariant under $\beta_n \rightarrow \beta_{4-n}$, $\alpha\leftrightarrow \beta$, $r \rightarrow r^{-1}$. They have the same structure as in singly-coupled bigravity \cite{Akrami:2012vf,Solomon:2014dua}: there is a single Friedmann equation sourced by $\rho$ and $r$, while $r$ evolves according to an \emph{algebraic} equation whose only time dependence comes from $\rho$. Notice that due to \cref{eq:quartic} one can write \cref{eq:Heff} in many different equivalent ways. It is therefore dangerous to directly identify the factors in front of $\rho$ in \cref{eq:Heff} as a time-varying gravitational constant, and the term proportional to $m^2$ as a dynamical dark energy component: both of these effects are present, but they cannot be straightforwardly separated from each other.

From \cref{eq:quartic}, we see that as $\rho\rightarrow\infty$ in the far past, either $r\rightarrow \beta/\alpha$ or $r\rightarrow -\alpha/\beta$. One can show that if $\rho\sim a^{-p}$ then $H^2\sim a^{-2p/3}$ as $r\rightarrow -\alpha/\beta$. Since this scenario is observationally excluded, we will not consider this limit.\footnote{In the singly-coupled theory there are ``infinite-branch" solutions where $r\to\infty$ at early times \cite{Konnig:2013gxa}. These infinite-branch solutions are crucial in the singly-coupled theory in order to avoid perturbative instabilities \cite{Solomon:2014dua,Konnig:2014xva}. However, in doubly-coupled bigravity, there are no solutions to \cref{eq:quartic} in which $r\rightarrow\infty$ as $\rho\rightarrow\infty$.}

An interesting feature is that in the early Universe
\begin{equation}
H^{2}\rightarrow\frac{(\alpha^2+\beta^2)\rho}{3M_{\eff}^{2}}.
\end{equation}
Since the coefficient in front of $\rho$ in the Friedmann equation during radiation domination can be probed by big bang nucleosynthesis, this could in principle be used to constrain the parameters of the theory. However, this will only work if the corresponding factor in front of $\rho$ in local gravity measurements has a different dependence on $\alpha$ and $\beta$. Whether this will be the case or not will be the subject of future work.

In the far future, as $\rho\rightarrow 0$, we have two possibilities. The first is that $r$ goes to a constant $r_c$, determined by
\begin{equation}\label{eq:quarticfuture}
\beta_{3}r_c^{4}+\left(3\beta_{2}-\beta_{4}\right)r_c^{3}+3\left(\beta_{1}-\beta_{3}\right)r_c^{2}+\left(\beta_{0}-3\beta_{2}\right)r_c-\beta_{1}=0.
\end{equation}
These models approach a de Sitter phase at late times (whether they \emph{self}-accelerate is a subtle question which we address below), with a cosmological constant given by
\begin{equation}\label{eq:effCC}
\Lambda=\frac{m^2\left[\beta_1+\left(\beta_0+3\beta_2\right)r_c+3\left(\beta_1+\beta_3\right)r_c^2+\left(3\beta_2+\beta_4\right)r_c^3+\beta_3 r_c^4\right]}{2r_c\left(\alpha+\beta r_c\right)^2}.
\end{equation}
The second possibility is that, for some parameter choices, $|r| \rightarrow\infty$ such that the leading-order $\beta_n$ term in \cref{eq:quartic} exactly cancels the leading density term, $r^4\rho$. It is unclear whether these solutions are viable; in this paper, we will restrict ourselves to solutions where $r$ is asymptotically constant in the past and future, starting at $r=\beta/\alpha$ and ending with $r=r_c$. This implies that $a_g$ and $a_f$ are proportional to one another in both the early and late Universe.

Furthermore, as long as $r$ does not exhibit any singular behavior, the evolution between $r=\beta/\alpha$ and $r=r_c$ is monotonic. This can be seen by taking a time derivative of \cref{eq:quartic} and setting $\dot r=0$. 

In the special case where $r_c=\beta/\alpha$, we see that $r=\beta/\alpha$ at all times, and the expansion history is identical to $\Lambda$CDM. This is a new feature of the doubly-coupled theory, as in the singly-coupled case, $r_c$ becomes zero in the presence of matter, which makes such a case trivially identical to general relativity. A constant $r$ occurs in any model where the $\beta_n$ parameters and $\beta/\alpha$ are chosen to satisfy
\begin{equation}\label{eq:betaalpha}
\beta_{3}\left(\frac{\beta}{\alpha}\right)^{4}+\left(3\beta_{2}-\beta_{4}\right)\left(\frac{\beta}{\alpha}\right)^{3}+3\left(\beta_{1}-\beta_{3}\right)\left(\frac{\beta}{\alpha}\right)^{2}+\left(\beta_{0}-3\beta_{2}\right)\left(\frac{\beta}{\alpha}\right)-\beta_{1}=0,
\end{equation}
which is simply \cref{eq:quarticfuture} with $r_c=\beta/\alpha$. An interesting implication of solutions with constant $r$ is that, since \cref{eq:NfNg} implies $N_f/N_g = da_f/da_g=r$, the two metrics are conformally related, $f_\mn = r^2 g_\mn$.\footnote{It is not difficult to see that there are no cases in which the two metrics are related by a \emph{dynamical} conformal factor; from \cref{eq:NfNg} any conformal relation means that $da_f/da_g=a_f/a_g$, but this implies $a_f/a_g=\mathrm{const.}$}

\section{Comparison to data: minimal models}
\label{sec:minimod}

In this section, we will compare the derived background expansion to observations and perform a parameter scan of the minimal models, in which only one of the $\beta_n$ is nonzero. Due to the duality property of the solutions, we only have to look at the $\beta_0$, $\beta_1$, and $\beta_2$ cases. We will restrict ourselves to positive $\beta/\alpha$; in principle negative values could also be allowed, but we have not yet investigated the physical implications of these values.\footnote{Note that $\beta<0$ leads to instabilities in the case of doubly-coupled dRGT massive gravity, in which one of the metrics is nondynamical \cite{Gumrukcuoglu:2014xba}.} The minimal models admit exact $\Lambda$CDM solutions when $\beta/\alpha=\left\{0,\frac{1}{\sqrt3},1\right\}$ for the $\beta_0$, $\beta_1$, and $\beta_2$ cases, respectively, as evident from \cref{eq:betaalpha}. 

Since we have so far calculated the equations of motion only for homogeneous backgrounds, we will limit this study to purely geometrical tests of the background expansion, 
including the redshift-luminosity relation of Type Ia supernovae (SNe) \cite{Suzuki:2011hu}, the observed angular scales of cosmic microwave background (CMB) 
anisotropies \cite{Ade:2013zuv}, and baryon-acoustic oscillations (BAO) \cite{Anderson:2012sa,Beutler:2011hx,Blake:2011en}. 
Since the latter two depend on the physical size of the sound horizon scale around the time when the CMB photons decoupled from the
baryon plasma, we can cancel out this dependence by using only the ratio of the observed angular scales in the CMB and BAO. In this way, we obtain a cosmological probe that is highly insensitive to the physics of the early Universe, and almost exclusively sensitive to the expansion history of the Universe between $z\sim 1000$ and today. The method was first used, and is extensively explained, in \cite{Sollerman:2009yu}. It was subsequently employed to constrain singly-coupled bigravity in \cite{vonStrauss:2011mq,Akrami:2012vf}. In the current paper, we use the most recent data as presented in the references given above.

We can calculate the effective equation of state for the background model described in eqs.~(\ref{eq:Heff}) and (\ref{eq:quartic}) using
\begin{equation}
w_\eff=-1-\frac{1}{3}\frac{d \log H^2}{d \log a}.
\end{equation}
Since in this paper we restrict ourselves to solutions where $r$ approaches constant values in the 
infinite past and future, for matter-dominated models we are guaranteed to have an 
effective equation of state where $w_\eff\rightarrow 0$ as $a\rightarrow 0$ (ignoring radiation) and $w_\eff\rightarrow -1$ as $a\rightarrow\infty$, mimicking the asymptotic behavior of the $\Lambda$CDM model. Except for some special parameter choices which are exactly $\Lambda$CDM (see the discussion above, as well as \cref{sec:specialparams}), we expect the model to deviate from the concordance model at all finite times.

It is well-known that $\Lambda$CDM is able to provide an excellent fit to background expansion data, so we expect the success of the bimetric model to depend on how close the effective equation of state is to that of $\Lambda$CDM. All solutions that look exactly like $\Lambda$CDM will trivially be able to fit existing background expansion data. Note, however, that this does not mean that these models are equivalent to $\Lambda$CDM, since they may give different predictions for perturbations, i.e., when studying structure formation.

\begin{figure}
\begin{centering}
\includegraphics[scale=0.4]{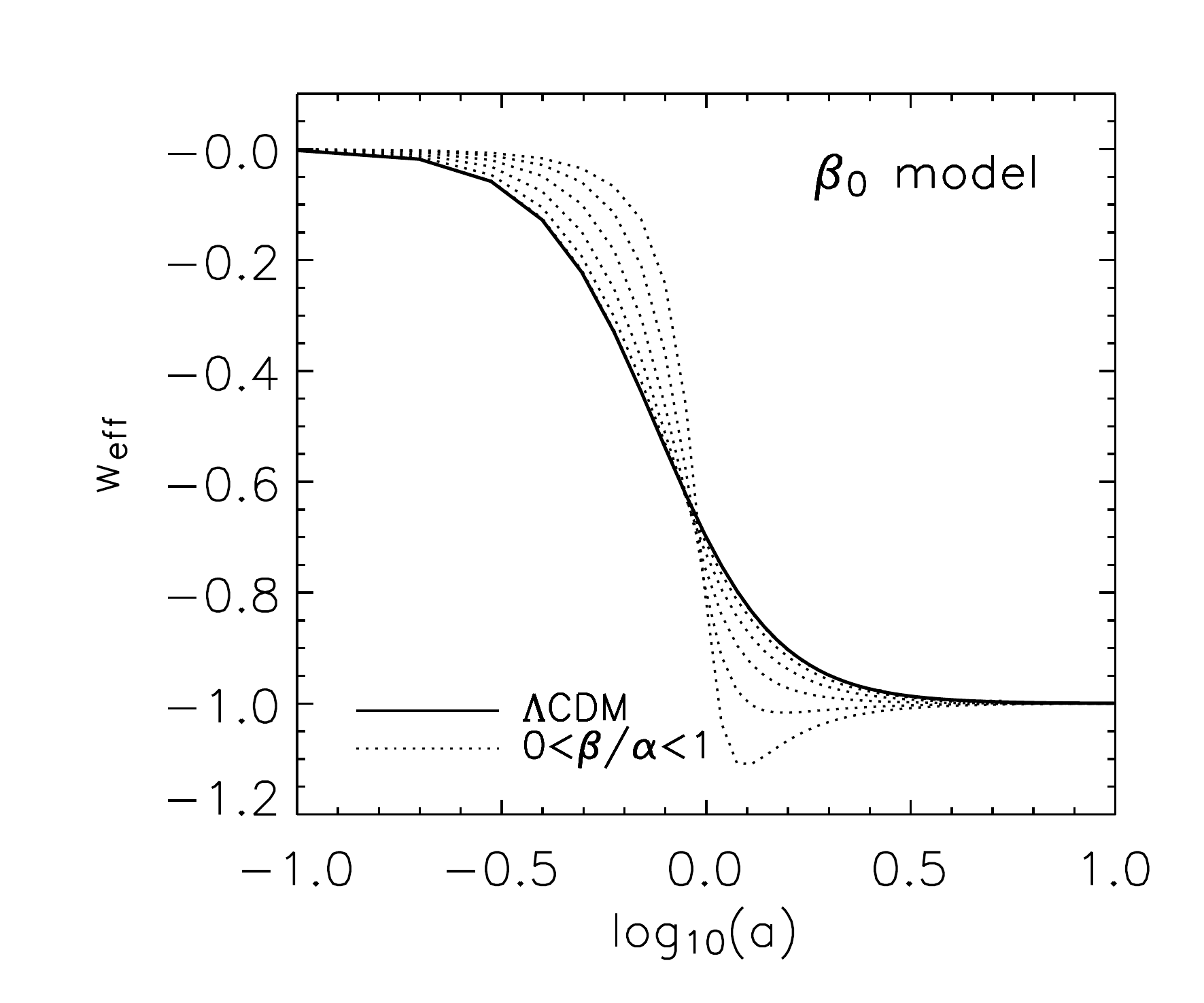}
\includegraphics[scale=0.4]{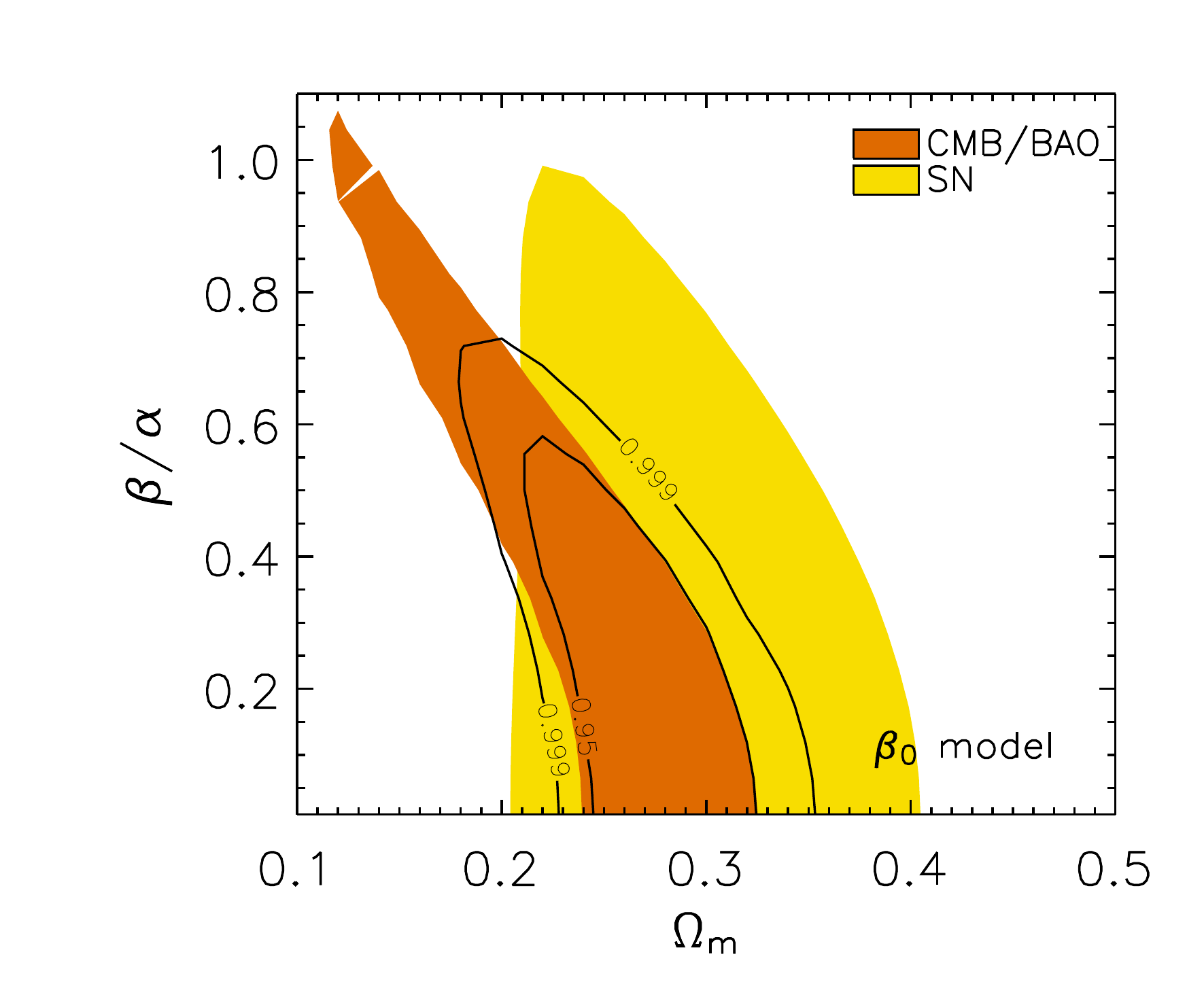}
\par\end{centering}
\caption{\label{fig:weffd0} {\em Left panel:} The effective equation of state, $w_\eff$, for the $\beta_0$ model with $0<\beta/\alpha<1$  (dotted lines) compared to $w_\eff$ of the $\Lambda$CDM model (solid line). When $\beta/\alpha\rightarrow 0$, the effective equation of state for the $\beta_0$ model approaches that of the $\Lambda$CDM model. In all cases, $\Omega_m=0.3$. {\em Right panel:} Confidence contours for $\Omega_m$ and $\beta/\alpha$ for the $\beta_0$ model as fitted to SNe, CMB, and BAO data.}
\end{figure}

In \cref{fig:weffd0}, we study the $\beta_0$ model, i.e., when only $\beta_0$ is turned on. Notice, cf. \cref{eq:Heff}, that this model has no nontrivial interactions between the two metrics, so it deviates from $\Lambda$CDM only through the novel matter coupling. In the left panel of \cref{fig:weffd0}, we compare the effective equation of state for different values of $\beta/\alpha$ with that of $\Lambda$CDM. We fix $\Omega_m=0.3$, where
\begin{equation}\label{eq:om}
\Omega_m \equiv \frac{\alpha^2\rho_0}{3M_\eff^2 H_0^2},
\end{equation}
and the subscript 0 indicates a value today. In the right panel of \cref{fig:weffd0}, we plot background constraints on $\Omega_m$ and $\beta/\alpha$. Note that the value of $\beta_0$ is set by the requirement that we have a flat geometry. Shaded contours show constraints from SNe and CMB/BAO data, respectively, corresponding to a 95\% confidence level for two parameters. Combined constraints are shown with solid lines corresponding to 95\% and 99.9\% confidence levels for two parameters.  As expected, when $\beta/\alpha \rightarrow 0$, the effective equation of state coincides with $\Lambda$CDM since this limit corresponds to the singly-coupled case where $\beta_0$ acts as a cosmological constant.  Note also that as $\beta/\alpha$ is increased, so is the factor multiplying the matter density in the Friedmann equation, and therefore the preferred matter density, $\Omega_m$, becomes smaller. This can be seen by setting $r=\beta/\alpha$ in \cref{eq:Heff} and writing the matter density in terms of $\Omega_m$ as defined in \cref{eq:om}, giving
\begin{equation}
\left(\frac{H}{H_0}\right)^{2}=\frac{\Omega_m}{a^3}\left[1+\left(\frac{\beta}{\alpha}\right)^2\right]+{\rm const.}
\end{equation}
Here, the last term, corresponding to the effective cosmological constant, takes the value $1-\Omega_m\left[1+\left(\beta/\alpha\right)^2\right]$ from the requirement of a flat spatial geometry.

\begin{figure}
\begin{centering}
\includegraphics[scale=0.4]{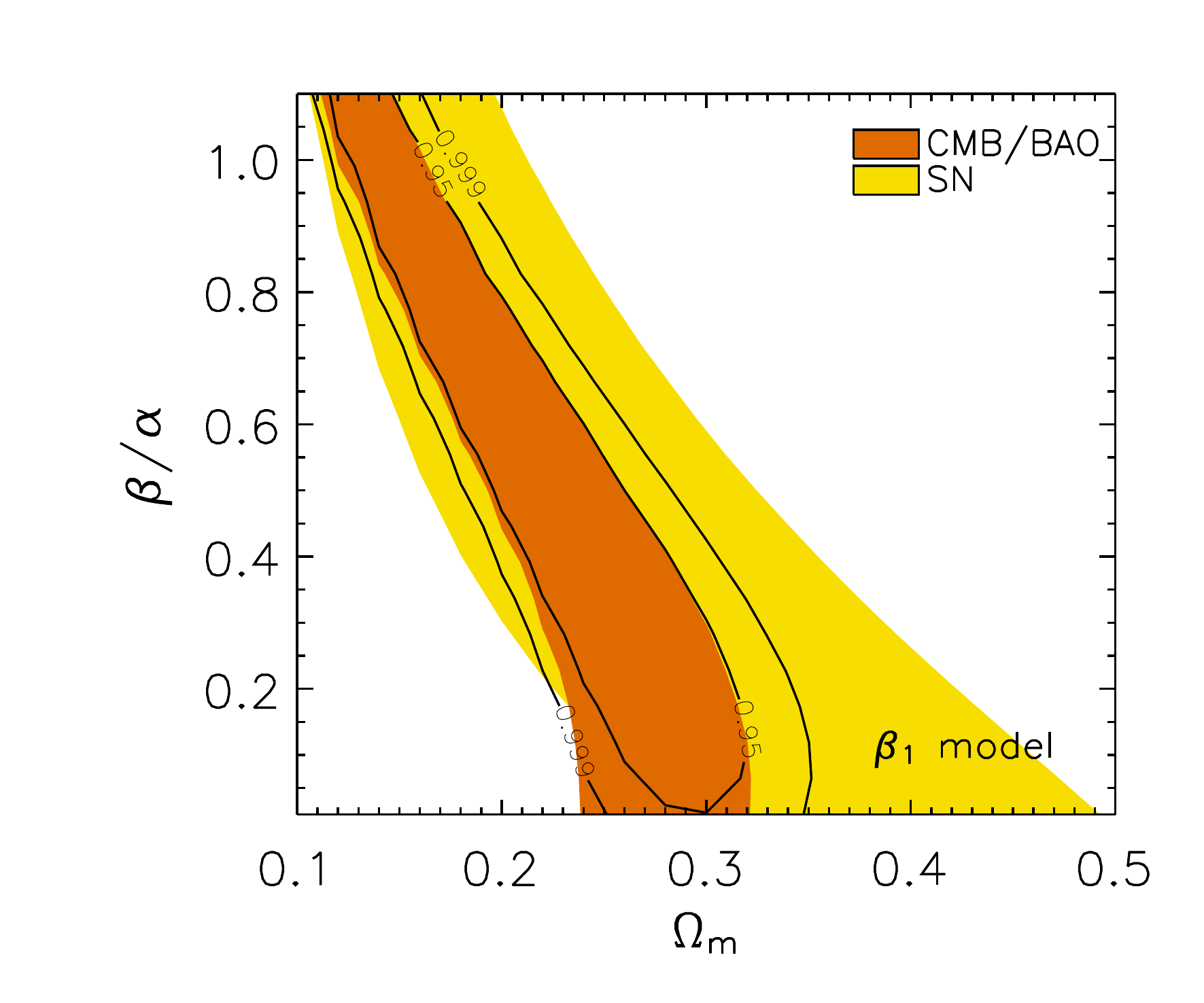}
\includegraphics[scale=0.4]{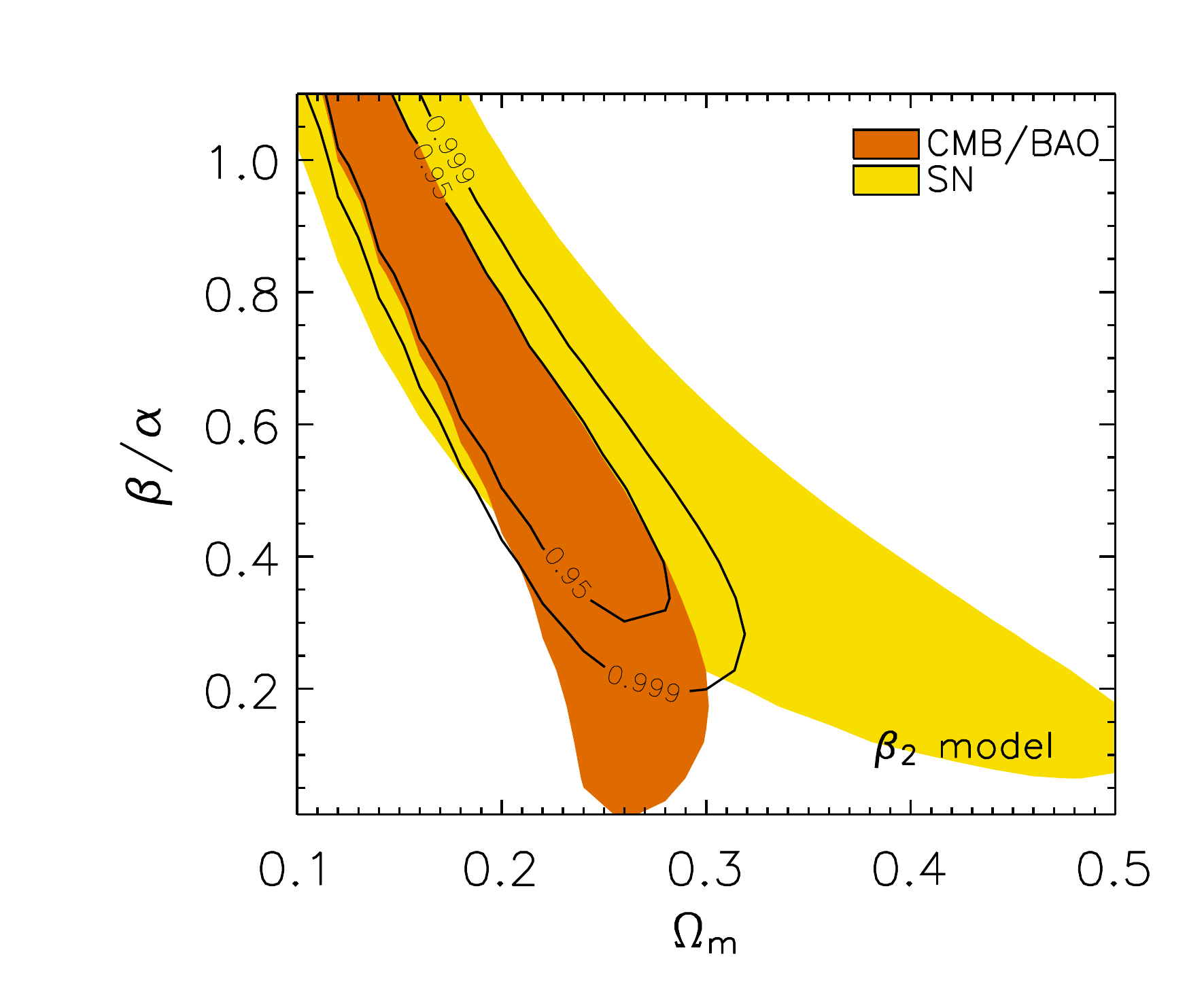}
\par\end{centering}
\caption{\label{fig:comb}Confidence contours for $\Omega_m$ and $\beta/\alpha$ for the $\beta_1$ and $\beta_2$ minimal models as fitted to SNe, CMB, and BAO data. In each case, we are able to obtain as good a fit as the concordance $\Lambda$CDM model.}
\end{figure}

In \cref{fig:comb} we plot background constraints on the $\beta_1$ and $\beta_2$ models. Since we know that the values $\beta/\alpha=\frac{1}{\sqrt{3}}$ and $\beta/\alpha=1$ give exact $\Lambda$CDM solutions for the $\beta_1$- and $\beta_2$-only models, respectively, we expect these values to provide good fits to the data. This is indeed the case, as can be seen in the plots.
The $\beta_2$ model is especially interesting in this regard, as $\beta/\alpha=1$ corresponds to the case where the two metrics $g_\mn$ and $f_\mn$ give equal contributions to the effective metric (or $M_g=M_f$ when using the $f$ and $g$ coupling strength framework described in appendix \ref{app:sym}). Notice that the $\beta_2$ model favors $\beta>0$, as we would expect since the $\beta_2$-only singly-coupled model is not in agreement with background data \cite{Akrami:2012vf} (this model is also ruled out by theoretical viability conditions \cite{Konnig:2013gxa}).

One of the attractive features of the double coupling is that it allows sensible cosmological solutions with only one of the $\beta_n$  turned on.
For more general combinations of the $\beta_n$ parameters, we expect the data to favor values that cluster around the value of $\beta/\alpha$ given by solving \cref{eq:betaalpha}, since this value yields an exact $\Lambda$CDM background expansion. We do not find it meaningful to do such a parameter scan at this moment, since it is only by including other probes, such as spherically symmetric solutions and cosmological perturbations, that we can exclude a larger part of the parameter space. However, in the next section, we discuss a few special cases that may turn out to be of particular interest for further investigations.

\section{Special parameter cases}
\label{sec:specialparams}

\subsection{Partially-massless gravity}
\label{sec:pm}

Partially-massless gravity is an interesting class of massive (bi)gravity theories where a new gauge symmetry might arise for the parameter choices \cite{deRham:2012kf,Hassan:2012gz}
\begin{equation}
\beta_0=3\beta_2=\beta_4, \qquad \label{eq:pmparams}
\beta_1=\beta_3=0.
\end{equation}
This gauge symmetry would eliminate the helicity-0 mode of the massive graviton, removing both the discontinuity between the $m\to0$ limit of linearized massive gravity and general relativity \cite{vanDam:1970vg,Zakharov:1970cc} and the fifth force which requires Vainshtein screening \cite{Vainshtein:1972sx} to reconcile the theory with solar system tests. Moreover, this new gauge symmetry would protect a small cosmological constant against quantum corrections. For more on partially-massless gravity, see \rcite{Hassan:2013pca}, as well as \rcite{Hassan:2012gz,Hassan:2014vja}, and references therein.

In singly-coupled bigravity, the partially-massless parameter choices could only be imposed in vacuum; including matter forces $r$ to be zero, which trivially reduces to general relativity. The nontrivial implications of the partially-massless scenario have been demonstrated for other doubly-coupled bigravity theories (see \rcite{Akrami:2013ffa}, though note that the theory discussed therein appears to have a ghost \cite{Yamashita:2014fga,deRham:2014naa}). Here we discuss this class in the context of the present doubly-coupled theory.

For the partially-massless parameter choices, \cref{eq:quartic} implies that $r=\beta/\alpha$, and the Friedmann equation becomes
\begin{align}
H^{2}=\frac{\alpha^2+\beta^2}{3M_\eff^{2}}\rho+\frac{m^{2}\beta_{0}}{3(\alpha^2+\beta^2)}.
\end{align}
This equation shows that the theory is equivalent to standard $\Lambda$CDM with an effective cosmological constant, $m^{2}\beta_0/(\alpha^2+\beta^2)$, and a rescaled gravitational coupling for matter. The partially-massless parameter choice will therefore give a background expansion that is identical to general relativity. Notice that this is a qualitatively new feature as compared to the singly-coupled theory.

Doubly-coupled bigravity with the parameters (\ref{eq:pmparams}) is thus a strong candidate partially-massless theory of gravity. In the context of single-metric (dRGT) massive gravity, with matter coupled only to the dynamical metric, this parameter choice leads to a theory which is not partially massless and in fact suffers from an infinitely strongly-coupled helicity-0 mode \cite{deRham:2013wv}. If doubly-coupled bigravity is shown to possess the partially-massless gauge symmetry nonlinearly and around all backgrounds, it should automatically become one of the most interesting available theories of gravity beyond general relativity.

\subsection{Vacuum energy and the question of self-acceleration}
\label{sec:vac}

One of the primary motivations for modifying general relativity is the possibility of having self-accelerating solutions, i.e., cosmologies which accelerate at late times even in the absence of a cosmological constant or vacuum energy contribution. In general relativity, as well as in singly-coupled bigravity, these two are degenerate: vacuum energy and a cosmological constant may have different origins, but they are mathematically indistinguishable. In bigravity with matter coupled to the effective metric, however, this question becomes rather subtle, as the vacuum energy from the matter sector produces more than just the cosmological constant terms for $g_\mn$ and $f_\mn$, which are equivalent to $\beta_0$ and $\beta_4$.

Matter loops will generate a term of the form $\sqrt{-\det g_\eff}\Lambda_v$. As shown in \rcite{deRham:2014naa}, the determinant of the effective metric is in fact a subset of the ghost-free interaction potential, with specific parameter choices for the $\beta_n$. This line of reasoning is partly what motivated \rcite{deRham:2014naa,Noller:2014sta} to consider this specific form of $g_\mn^\eff$ in the first place. In particular, the determinant has the property that
\begin{equation}
\sqrt{-\det g_\eff}=\sqrt{-\det g}\det\left(\alpha+\beta X\right).
\end{equation}
Since
\begin{equation}
\det\left(\alpha+\beta X\right)=\sum_{n=0}^{4}\alpha^{4-n}\beta^n e_{n}\left(X\right),
\end{equation}
we see that a pure vacuum energy contribution can be written in the form of the bigravity interaction potential with parameters
\begin{equation}\label{eq:lambdabeta}
\beta_n = \frac{\Lambda_v \alpha^{4-n}\beta^n}{m^{2}}.
\end{equation}

Let us assume that the $\beta_n$ parameters take this particular form, i.e., the only metric interactions arise from matter loops. The quartic equation~(\ref{eq:quartic}) can then be solved only if $r=\beta/\alpha$ (or $\rho = -M_\eff^2\Lambda_v$), and the Friedmann equation becomes
\begin{equation}\label{eq:lambdahubble}
H^{2}=\frac{\left(\alpha^{2}+\beta^{2}\right)\rho}{M_{\eff}^{2}}+\frac{\left(\alpha^{2}+\beta^{2}\right)\Lambda_v}{3}.
\end{equation}
\Cref{eq:lambdabeta,eq:lambdahubble} reduce to the known expression for the $\Lambda$CDM solutions with a cosmological constant proportional to either $\beta_0$ or $\beta_4$ in the singly-coupled limit (where either $\beta\rightarrow0$ or $\alpha\rightarrow0$).

It is of course not surprising that matter loops lead to an accelerating expansion. However, the appearance of the vacuum energy in \emph{all} the bigravity interaction terms has novel implications. First, because the vacuum energy contributes to all the interaction terms, the mass scale $m$ is not protected against quantum corrections from matter loops \cite{deRham:2014naa}. Therefore, any values we obtain for these parameters from comparison of the theory to observations must be highly fine-tuned.\footnote{If the case described in \cref{sec:pm} is truly partially massless, this may be an exception, as there is a new gauge symmetry to protect against quantum corrections.} This is in contrast to singly-coupled bigravity, in which the only parameter that receives contributions from quantum loops is $\beta_0$ (if one couples matter to $g_\mn$), just as in general relativity where the cosmological constant is unstable in the presence of matter fields. In the singly-coupled theory, the scale $m$ and the structure of the interaction potential are stable to quantum corrections \cite{deRham:2012ew,deRham:2013qqa}, a very useful fact which is lost once we couple matter to $g^\eff_\mn$.

The other implication is that self-accelerating solutions are no longer straightforward to define in this theory. Typically, self-acceleration refers to cosmologies which accelerate at late times even when the vacuum energy is set to zero. Since in general relativity and singly-coupled bigravity, there is a single parameter which is degenerate with the vacuum energy ($\Lambda$ in the former and $\beta_0$ or $\beta_4$ in the latter), one can simply set its value to zero and look for other accelerating solutions. In the present doubly-coupled theory, however, \emph{all} interaction terms are degenerate with the vacuum energy: given an interaction potential, there is no way to unambiguously determine the value of $\Lambda_v$. In that respect, we cannot set some of the parameters to zero in order to restrict ourselves to accelerating solutions arising from nonvacuum, massive-gravity interaction terms (unless we set all the parameters to zero, which will give uninteresting solutions). Therefore, from a particle physics point of view our theory lacks self-accelerating solutions.

\subsection{Maximally-symmetric model}

The parameter values $\beta_0=\beta_4$, $\beta_1=\beta_3$, and $\alpha=\beta$ are special in the sense that they map a solution to itself under the transformation $g_\mn\leftrightarrow f_\mn$, $\beta_n \rightarrow \beta_{4-n}$, $\alpha \leftrightarrow \beta$ (vacuum solutions for this model were previously studied in \rcite{Hassan:2014vja}). Thus this theory is maximally symmetric between the two metrics; they appear in the theory in completely equal ways. Eq.~(\ref{eq:quartic}) becomes
\begin{equation}
\left(r^{2}-1\right)\left[\beta_{1}\left(r^{2}+1\right)+3\beta_{2}r-\beta_{0}r+\frac{\rho\alpha^{4}}{m^{2}M_{\eff}^{2}}\left(1+r\right)^{2}\right]=0.
\end{equation}
The exact $\Lambda$CDM solution in this case is, as expected, given by $r=1$. (Indeed, the two metrics are completely equal, $g_\mn = f_\mn$, due to the Bianchi constraint $N_f/N_g=da_f/da_g=1$.) The second order polynomial for $r$ in brackets gives two solutions, which are inverses of one another. This is expected, since when $g_\mn\leftrightarrow f_\mn$ we have $r\rightarrow r^{-1}$.

\section{Conclusions}
\label{sec:Conclusions}

In this paper we have presented the main features of the background expansion for massive bigravity with matter ``doubly coupled" to both metrics through an effective metric, given by
\begin{equation}
g^\eff_{\mu\nu} = \alpha^2 g_{\mu\nu} + 2\alpha\beta g_{\mu\alpha}X^{\alpha}_{\nu}+\beta^2 f_{\mu\nu},\qquad X^{\mu}_{\nu} = (\sqrt{g^{-1}f})^\mu_\nu.
\end{equation}
This coupling was introduced in \rcite{deRham:2014naa,Noller:2014sta}, and has been further discussed in \rcite{deRham:2014fha,Hassan:2014gta,Noller:2014ioa,Solomon:2014iwa,Gumrukcuoglu:2014xba}. The expansion history is described by \cref{eq:Heff,eq:quartic},
\begin{align}
H^{2}&=\frac{\rho}{6M_{\eff}^{2}}\left(\alpha+\beta r\right)\left(\alpha+\beta r^{-1}\right)+\frac{m^{2}\left(B_{0}+r^{2}B_{1}\right)}{6\left(\alpha+\beta r\right)^{2}}, \\
0 &= \frac{\rho}{M_\eff^{2}}\left(\alpha+\beta r\right)^{3}\left(\alpha -\beta r^{-1}\right)+m^{2}\left(B_0-r^2B_1\right). \label{eq:quarticconc}
\end{align}
where $B_0$ and $B_1$ are defined in \cref{eq:B0,eq:B1}. The first of these is the Friedmann equation for the effective metric, and the second algebraically describes the evolution of $r=a_f/a_g$, the ratio of the $f$- and $g$-metric scale factors. One can always choose the parameters of the theory such that the background expansion is exactly that of $\Lambda$CDM; any parameter choice which leads to $r=\beta/\alpha$ in \cref{eq:quarticconc} will have this behavior. For more general parameter values, the background expansion will deviate from $\Lambda$CDM, but may still be consistent with observational data. In section \ref{sec:minimod} we place constraints on the models with only $\beta_0$, $\beta_1$, or $\beta_2$ nonzero. The other single-parameter models---with $\beta_3$ or $\beta_4$ nonzero---are then automatically included in this analysis due to the duality between solutions under $g_\mn\leftrightarrow f_\mn$, $\beta_n \rightarrow \beta_{4-n}$, and $\alpha\leftrightarrow\beta$, as described in \cref{app:sym}.

A novel feature of the effective coupling studied here is that $g_\mn$ and $f_\mn$ can be conformally related in the presence of matter at the background level. In the singly-coupled case, this is only possible in vacuum, where the solutions are de Sitter. A special case is the parameter choice leading to a candidate partially-massless theory, a theory that potentially has a novel gauge symmetry which would eliminate the problematic fifth force and protect a small vacuum energy against quantum corrections. In this case the background is identical to $\Lambda$CDM in the presence of matter. This suggests that doubly-coupled bigravity is a promising candidate for a theory of partially-massless gravity.

Concerning the perturbations, since singly-coupled bigravity tends to be unstable for small $r$ \cite{Konnig:2014xva}, there is hope that these doubly-coupled models will have better stability properties as $r$ is always nonzero and can be made to have an arbitrarily large minimum value.

On the whole, the matter coupling studied in this paper has several advantages. Notably, it retains the metric interchange symmetry and has a straightforward physical interpretation. Since this coupling is phenomenologically viable when it comes to the background expansion, we belive that other phenomenological studies are well-motivated. 

\acknowledgments

We would like to thank Fawad Hassan, Mikica Kocic, Tomi Koivisto, Frank K\"{o}nnig, Malin Renneby and Angnis Schmidt-May for many fruitful discussions in relation to this work. A.R.S. is supported by the David Gledhill Research Studentship, Sidney Sussex College, University of Cambridge; and by the Isaac Newton Fund and Studentships, University of Cambridge. Y.A. is supported by the European Research Council (ERC) Starting Grant StG2010-257080. E.M. acknowledges support for this study by the Swedish Research Council.

\appendix

\section{Transformation properties of the action}
\label{app:sym}

Here we describe the transformation properties of the action and how they determine the number of physically-relevant parameters.

\subsection{Rescaling the action}
We write the action as
\begin{align}
S & =  -\frac{M_{g}^{2}}{2}\int d^{4}x\sqrt{-\det g}R\left(g\right)-\frac{M_{f}^{2}}{2}\int d^{4}x\sqrt{-\det f}R\left(f\right) \nonumber \\
 &  \hphantom{{}=}+m^{4}\int d^{4}x\sqrt{-\det g}\mathcal{V}\left(\sqrt{g^{-1}f};\beta_n\right) +\int d^{4}x\sqrt{-\det g_\eff}\mathcal{L}_{m}\left(g_\eff,\Phi\right),
\end{align}
where
\begin{equation}
\mathcal{V}\left(\sqrt{g^{-1}f};\beta_n \right) = \sum_{n=0}^{4}\beta_{n}e_{n}\left(\sqrt{g^{-1}f}\right)
\end{equation}
is the interaction potential, which satisfies
\begin{equation}
\sqrt{-\det g}\mathcal{V}\left(\sqrt{g^{-1}f};\beta_{n}\right)=\sqrt{-\det f}\mathcal{V}\left(\sqrt{f^{-1}g};\beta_{4-n}\right).
\end{equation}
Due to this property, the action is invariant under
\begin{equation}
g_\mn\leftrightarrow f_\mn, \qquad M_g \leftrightarrow M_f, \qquad \alpha \leftrightarrow \beta, \qquad \beta_n \rightarrow \beta_{4-n}, 
\end{equation}
since the effective metric
\begin{equation}
g^\eff_\mn = \alpha^2 g_\mn +2\alpha\beta g_{\mu\alpha} X^\alpha_\nu + \beta^2 f_\mn
\label{eq:appgeff}
\end{equation}
is also invariant under this transformation, as shown in \cref{app:sym1}. Because the overall scaling of the action is unimportant, there is a related transformation which keeps the action invariant, but only involves the ratio of $M_g$ and $M_f$: 
\begin{equation}
\frac{M_{g}}{M_{f}}g_{\mn}\leftrightarrow\frac{M_{f}}{M_{g}}f_{\mn},\qquad\left(\frac{M_{f}}{M_{g}}\right)^{4-n}\beta_{n}\rightarrow\left(\frac{M_{f}}{M_{g}}\right)^{n}\beta_{4-n},\qquad\frac{M_{f}}{M_{g}}\alpha^{2}\leftrightarrow\frac{M_{g}}{M_{f}}\beta^{2}.
 \end{equation}
These transformations reflect a \textit{duality} of the action since they map one set of solutions, with a given set of parameters, to another set of solutions.

Not all of the parameters $M_g$, $M_f$, $\alpha$, $\beta$, and $\beta_n$ are physically independent. In effect, we can rescale these parameters, together with $g_\mn$ and $f_\mn$, to get rid of either $M_g$ and $M_f$ \textit{or} $\alpha$ and $\beta$. In the end, only the ratio between $M_g$ and $M_f$, or $\alpha$ and $\beta$, together with $\beta_n$, are physically meaningful. The two parameter choices are physically equivalent and can be mapped to one another. We now describe the two scalings that give rise to the two parameter choices.

Under the scalings
\begin{align}
&& g_{\mu\nu}&\rightarrow\alpha^{-2}g_{\mu\nu}, & f_{\mu\nu}&\rightarrow\beta^{-2}f_{\mu\nu}, & M_{g}^{2}&\rightarrow\alpha^{2}M_{g}^{2}, && \nonumber \\
&& M_{f}^{2}&\rightarrow\beta^{2}M_{f}^{2}, &  m^{4}&\rightarrow\alpha^{4}m^{4}, & \beta_{n}&\rightarrow\left(\frac{\beta}{\alpha}\right)^{n}\beta_{n}, &&
\end{align}
the effective metric becomes
\begin{equation}
g^\eff_\mn = g_\mn + 2 g_{\mu\alpha} X^\alpha_\nu + f_\mn
\end{equation}
while the action becomes
\begin{align}
S & =  -\frac{M_{g}^{2}}{2}\int d^{4}x\sqrt{-\det g}R\left(g\right)-\frac{M_{f}^{2}}{2}\int d^{4}x\sqrt{-\det f}R\left(f\right) \nonumber \\
 &  \hphantom{{}=}+m^{4}\int d^{4}x\sqrt{-\det g}\sum_{n=0}^{4}\beta_{n}e_{n}\left(\sqrt{g^{-1}f}\right) \nonumber \\
 &  \hphantom{{}=} +\int d^{4}x\sqrt{-\det g_\eff}\mathcal{L}_{m}\left(g_\eff,\Phi\right).
\end{align}
The effective metric is thus uniquely defined in this parameter framework, whereas the ratio between $M_g$ and $M_f$ is the free parameter besides the $\beta_n$. For this choice of scaling, the action is invariant under
\begin{equation}
g_{\mu\nu}\leftrightarrow f_{\mu\nu},\quad\beta_{n}\rightarrow\beta_{4-n},\qquad M_{g}\leftrightarrow M_{f},
\end{equation}
or, more generally (as described in \rcite{Hassan:2014vja}),
\begin{equation}
\frac{M_{g}}{M_{f}}g_{\mu\nu}\leftrightarrow\frac{M_{f}}{M_{g}}f_{\mu\nu},\qquad\left(\frac{M_{f}}{M_{g}}\right)^{4-n}\beta_{n}\rightarrow\left(\frac{M_{f}}{M_{g}}\right)^{n}\beta_{4-n}.
\end{equation}

If, instead, we apply the scalings
\begin{align}
&& g_{\mu\nu}&\rightarrow\frac{M_\eff^{2}}{M_{g}^{2}}g_{\mu\nu},& f_{\mu\nu}&\rightarrow\frac{M_\eff^{2}}{M_{f}^{2}}f_{\mu\nu}, & \beta_{n}&\rightarrow\left(\frac{M_{f}}{M_{g}}\right)^{n}\beta_{n},&& \nonumber \\
&& m^{4}&\rightarrow m^{2}\frac{M_{g}^{4}}{M_\eff^{2}}, & \alpha^{2}&\rightarrow\frac{M_{g}^{2}}{M_\eff^{2}}\alpha^{2},&\beta^{2}&\rightarrow\frac{M_{f}^{2}}{M_\eff^{2}}\beta^{2},&&
\label{eq:Meffscaling}
\end{align}
the effective metric is still of the form given in \cref{eq:appgeff}, while the action becomes
\begin{align}
S & =  -\frac{M_\eff^{2}}{2}\int d^{4}x\sqrt{-\det g}R\left(g\right)-\frac{M_\eff^{2}}{2}\int d^{4}x\sqrt{-\det f}R\left(f\right) \nonumber \\
 &  \hphantom{{}=}+m^{2}M_\eff^2 \int d^{4}x\sqrt{-\det g}\sum_{n=0}^{4}\beta_{n}e_{n}\left(\sqrt{g^{-1}f}\right) \nonumber \\
 &  \hphantom{{}=} +\int d^{4}x\sqrt{-\det g_\eff}\mathcal{L}_{m}\left(g_\eff,\Phi\right).
\end{align}
For this choice of scaling, only the ratio between $\alpha$ and $\beta$, together with the $\beta_n$, is physically important (the effective coupling $M_\eff^2$ can be absorbed in the normalization of the matter content). Under this form, the action is invariant under
\begin{equation}
g_{\mu\nu}\leftrightarrow f_{\mu\nu},\qquad\beta_{n}\rightarrow\beta_{4-n},\qquad\alpha\leftrightarrow\beta.
\end{equation}

To move from the framework with $M_g$ and $M_f$ to that of $\alpha$ and $\beta$, one simply performs the rescaling
\begin{align}
M_g^2&\rightarrow \frac{M_\eff^2}{\alpha^2}, \qquad M_f^2 \rightarrow \frac{M_\eff^2}{\beta^2}, \qquad g_\mn \rightarrow \alpha^2 g_\mn, \nonumber \\
f_\mn&\rightarrow \beta^2 f_\mn, \qquad \beta_n \rightarrow \left(\frac{\alpha}{\beta}\right)^n \beta_n, \qquad m^4\rightarrow \frac{m^2M_\eff^2}{\alpha^4}.
\end{align}

The different parameter frameworks have their respective advantages: In the $M_g$ and $M_f$ framework, there is a {\it unique} effective metric, and it is the relative coupling strengths that determine the physics. In the $\alpha$ and $\beta$ framework, we have one gravitational coupling strength $M_\eff^2$, and the singly-coupled limits are more apparent. Note also that the ratio between $\alpha$ and $\beta$ only appears in the matter sector, whereas in the $M_g$ and $M_f$ formulation their ratio appears in both the matter sector and interaction potential.  

\subsection{Symmetry of the effective metric}
\label{app:sym1}

In this section, we show that the effective metric is symmetric under the interchange $g_\mn \leftrightarrow f_\mn$, $\alpha \leftrightarrow \beta$. In order to do this, we take advantage of the fact that $g_{\mu\alpha} X^\alpha_\nu$ is symmetric, i.e., $g X = X^T g$, as shown in \rcite{Hassan:2012wr}. We will find it useful to discuss the metrics in terms of their vielbeins, since we are dealing with square root matrices and vielbeins are, in a sense, ``square roots" of their respective metrics. We use Greek letters for spacetime indices and Latin letters for Lorentz indices. The $g$ and $f$ metric vielbeins are defined by
\begin{align}
 g_\mn &\equiv \eta_\ab e^a_\mu e^b_\nu, \\
 f_\mn &\equiv \eta_\ab L^a_\mu L^b_\nu,
\end{align}
while the inverse metrics are given by $g^\mn = \eta^\ab e_a^\mu e_b^\nu$ and similarly for $f^\mn$. The vielbeins of $g^\mn$ are inverses of the vielbeins for $g_\mn$, $e^a_\mu e_b^\mu = \delta^a_b$ and $e^a_\mu e^\nu_a = \delta^\nu_\mu$, and again similarly for the $f_\mn$ vielbeins.

We will assume the symmetry condition (also called the Deser-Van Nieuwenhuizen gauge condition)
\begin{equation}
e_a^\mu L_{b \mu} = e^\mu_bL_{a \mu}, \label{eq:symmcond} 
\end{equation}
where Lorentz indices are raised and lowered with the Minkowski metric. It is likely, though it has not yet been proven, that this condition holds for all physically relevant cases. In four dimensions, it holds when $g^{-1}f$ has a real square root (this was proven in \rcite{Deffayet:2012zc}, where it was conjectured that this result is valid also in higher dimensions). Assuming this condition, then it has been shown \cite{Gratia:2013gka} that the square root matrix is given by
\begin{equation}
 X^\mu_\nu = e^\mu_aL^a_\nu. \label{eq:sqrtviel}
\end{equation}
The inverse of this is clearly
\begin{equation}
(X^{-1})_{\nu}^{\mu}=L_{a}^{\mu}e_{\nu}^{a},
\end{equation}
since
\begin{equation}
X_{\alpha}^{\mu}(X^{-1})_{\nu}^{\alpha}=e_{a}^{\mu}L_{\alpha}^{a}L_{b}^{\alpha}e_{\nu}^{b}=e_{a}^{\mu}e_{\nu}^{a}=\delta_{\nu}^{\mu}.
\end{equation}
The form of the inverse then implies
\begin{equation}
\left(\sqrt{f^{-1}g}\right)^{-1}=\sqrt{g^{-1}f}, \label{eq:Xinv}
\end{equation}
which will be a useful property when showing the symmetry of the effective metric. We also have
\begin{equation}
g_{\mu\alpha}X_{\nu}^{\alpha}=e_{a\mu}e_{\alpha}^{a}e_{b}^{\alpha}L_{\nu}^{b}=e_{a\mu}L_{\nu}^{a}.
\end{equation}
In order to show that $g X = X^T g$, we must thus have
\begin{equation}
e_{a\mu}L_{\nu}^{a}=e_{a\nu}L_{\mu}^{a}.
\end{equation}
Notice that this is not exactly the same as eq.~(\ref{eq:symmcond}),
since in the first case we contract over spacetime indices, whereas
here we have contracted over Lorentz indices. The two symmetry conditions
are, however, equivalent, as discussed in detail in \rcite{Hoek:1982za}. An alternative way of seeing that $gX=X^Tg$ is as follows. Since
\begin{equation}
f_{\mu\alpha}X_{\nu}^{\alpha}=L_{\mu}^{a}L_{a\alpha}e_{b}^{\alpha}L_{\nu}^{b}=L_{\mu}^{a}L_{b\alpha}e_{a}^{\alpha}L_{\nu}^{b}=f_{\nu\alpha}X_{\mu}^{\alpha},
\end{equation}
we have
\begin{equation}
fX=X^{T}f.
\label{eq:fXsym}
\end{equation}
But $f = gX^2$, so \cref{eq:fXsym} can also be written $gX^{3}=X^{T}gX^{2}$, which implies
\begin{equation}
gX=X^{T}g.
\label{eq:gXsym}
\end{equation}

Using this property it is straightforward to show that the effective metric is symmetric under the interchange of the two metrics. The effective metric we study was introduced in \rcite{deRham:2014naa} in the form
\begin{equation}
g_{\mu\nu}^\eff=\alpha^{2}g_{\mu\nu}+2\alpha\beta g_{\alpha(\mu}X_{\nu)}^{\alpha}+\beta^{2}f_{\mu\nu}.
\end{equation}
Due to the symmetry property (\ref{eq:gXsym}), we can write this without the explicit symmetrization,
\begin{equation}
g_{\mu\nu}^\eff=\alpha^{2}g_{\mu\nu}+2\alpha\beta g_{\alpha\mu}X_{\nu}^{\alpha}+\beta^{2}f_{\mu\nu}.
\end{equation}
Suppose now that we do the transformation
\begin{equation}
g_\mn\leftrightarrow f_\mn,\qquad\alpha\leftrightarrow\beta.
\end{equation}
The effective metric becomes
\begin{equation}
\label{eq:gefftrans}
g_{\mu\nu}^\eff=\alpha^{2}g_{\mu\nu}+2\alpha\beta f_{\mu\alpha}(\sqrt{f^{-1}g})_{\nu}^{\alpha}+\beta^{2}f_{\mu\nu}.
\end{equation}
However,
\begin{equation}
f\left(\sqrt{g^{-1}f}\right)^{-1}=gg^{-1}f\left(\sqrt{g^{-1}f}\right)^{-1}=g\sqrt{g^{-1}f}.
\end{equation}
(This property was used in the original construction of the interaction potential in \rcite{Hassan:2011vm}.)
Combining this with \cref{eq:Xinv} we get
\begin{equation}
f\sqrt{f^{-1}g}=g\sqrt{g^{-1}f}.
\end{equation}
Applying this to \cref{eq:gefftrans} we see that the effective metric is invariant under the duality transformation. This ensures that the entire Hassan-Rosen action treats the two metrics on entirely equal footing when using the effective coupling. Note that this duality does not hold for the single-metric (dRGT) massive gravity with the effective metric, as it is broken in the kinetic sector.

\section{Deriving the equations of motion}
\label{app:eom}

In this appendix we derive the cosmological equations by directly varying the Lagrangian, rather than using the full Einstein equations. The Einstein equations with this matter coupling are highly nontrivial to derive \cite{Schmidt-May:2014xla}, so this method provides a good check. We employ the scaling given in \cref{eq:Meffscaling} to replace $M_g$ and $M_f$ with a single coupling $M_\eff$. The gravitational Lagrangian in an FLRW universe is
\begin{equation}
\calL=3\frac{a_g\dot{a}_g^{2}}{N_g}+3\frac{a_f\dot{a}_f^{2}}{N_f}+m^{2}N_ga^{3}_g\sum_{n=0}^{4}\beta_{n}e_{n}+\left(\alpha N_g+\beta N_f\right)\left(\alpha a_g+\beta a_f\right)^{3}\frac{\rho(a)}{M_\eff^{2}}.
\end{equation}
The equations of motion are given by varying the Lagrangian with respect to $N_g$, $a_g$, $N_f$, and $a_f$. We define the pressure by
\begin{equation}
\label{eq:pressure}
p(a)=-\frac{1}{3}\left(3\rho(a)+a\frac{\delta\rho(a)}{\delta a}\right).
\end{equation}
Using the definitions presented in \cref{sec:flrw}, together with
\begin{equation}
P\equiv \beta_1 + 2\beta_2 r +\beta_3 r^2,
\end{equation}
the equations of motion are
\begin{align}
3\frac{a_g\dot{a}_g^{2}}{N_g^{2}}-m^{2}a_g^{3}B_{0}&=\frac{\alpha}{M_\eff^{2}}a^{3}\rho(a), \\
\frac{\dot{a}_g^{2}}{N_g}-\frac{d}{dt}\left(\frac{2a_g\dot{a}_g}{N_g}\right)+m^{2}N_ga_g^{2}\left[B_{0}+\left(\frac{N_f}{N_g}-\frac{a_f}{a_g}\right)P\right]&=\frac{\alpha}{M_\eff^{2}}N a^{2}p(a), \\
3\frac{a_f\dot{a}_f^{2}}{N_f^{2}}-m^{2}a_f^{3}B_{1}&=\frac{\beta}{M_\eff^{2}}a^{3}\rho(a), \\
\frac{\dot{a}_f^{2}}{N_f}-\frac{d}{dt}\left(\frac{2a_f\dot{a}_f}{N_f}\right)+m^{2}\frac{N_f}{a_f}\left[a_f^3B_{1}+Pa_g^{3}\left(\frac{a_fN_g}{a_gN_f}-1\right)\right]&=\frac{\beta}{M_\eff^{2}}N a^{2}p(a).
\end{align}
 Using
\begin{equation}
\dot{B}_{0}=3P\dot r,\qquad \dot{B}_{1}=-3P\frac{\dot r}{r^4},
\end{equation}
with $r\equiv a_f/a_g$, we can derive the following conditions from the equations of motion:
\begin{align}
\dot{\rho}+3\frac{\dot{a}}{a}\left(\rho+\frac{N}{N_g}\frac{\dot{a}_g}{\dot{a}}p\right)&=-\frac{3m^{2}M_\eff^{2}a_g^{2}P}{\alpha a^{3}N_g}\left(N_g\dot{a}_f-\dot{a}_gN_f\right), \\
\dot{\rho}+3\frac{\dot{a}}{a}\left(\rho+\frac{N}{N_f}\frac{\dot{a}_f}{\dot{a}}p\right)&=\frac{3m^{2}M_\eff^{2}a_g^{2}P}{\beta a^{3}N_f}\left(N_g\dot{a}_f-\dot{a}_gN_f\right).
\end{align}
These two equations imply
\begin{equation}
\dot{\rho}+3\frac{\dot{a}}{a}\left(\rho+p\right)=0
\end{equation}
and
\begin{equation}
\left(m^{2}M_\eff^{2}a_g^{2}P - \alpha\beta a^{2}p\right)\left(N_f\dot{a}_g-N_g\dot{a}_f\right)=0.
\end{equation}
Besides being a consequence of the equations of motion, the continuity equation for $\rho$ is also implied in our definition of pressure (\ref{eq:pressure}).

\section{Algebraic branch}
\label{sec:firstbranch}

In this section, without any ambition to examine all possible solutions, we briefly outline some of the properties of a few specific solutions on the algebraic branch of the Bianchi constraint (\ref{eq:bianchi}). Setting the first bracket of \cref{eq:bianchi} to zero, we get
\begin{equation}
m^{2}\left(\beta_1 a_g^2+2\beta_2 a_g a_f + \beta_3 a_f^2\right)=\frac{\alpha\beta a^{2}p}{M_\eff^2}.
\label{eq:firstbrancheq}
\end{equation}
If the Universe is dominated by dust ($p=0$), this is a polynomial equation for $r$:
\begin{equation}
\beta_1+2\beta_2 r+\beta_3 r^2=0,
\end{equation}
which is solved by a constant $r=r_c$. Because $r$ is constant, the mass terms in the two Friedmann equations become constant, so $H_g$ and $H_f$ are determined by Friedmann equations containing effective cosmological constants.

These do not, however, necessarily lead to a $\Lambda$CDM cosmology for the effective metric. Using the fact that $a=\left(\alpha+\beta r_c\right)a_g=\left(\alpha/r_c+\beta\right)a_f$, we can show that the observed Hubble rate, $H = \dot a/(aN)$, for a constant $r$ is given by
\begin{equation}
H = H_g\left(\alpha + \beta\frac{N_f}{N_g}\right)^{-1} = H_f\left(\alpha\frac{N_g}{N_f} + \beta\right)^{-1}.
\end{equation}
We see that if the ratio $N_f/N_g$ is constant, the solutions on this branch contain an exact cosmological constant (at least at the background level) given by a combination of the metric interaction terms. If that ratio varies, then the observed Hubble rate will deviate from $\Lambda$CDM. We can write the ratio of the lapses as
\begin{equation}
\left(\frac{N_f}{N_g}\right)^2 = \frac{3\alpha H_g^2 r_c^3}{3\beta H_g^2+m^2\left(\alpha B_1 r_c^3 - \beta B_0\right)}, \label{eq:lapseratio}
\end{equation}
from which we see that the parameter choice $\alpha r_c^3 B_1=\beta B_0$ gives a constant $N_f/N_g$, and thus leads to $\Lambda$CDM behavior for the effective Hubble rate. Other parameter choices will lead to novel cosmological behavior.

For nonzero pressure, $p\neq0$, we can rewrite the constraint (\ref{eq:firstbrancheq}) as
\begin{equation}
m^{2}M_\eff^2\left(\beta_{1}+2\beta_{2}r+\beta_{3}r^{2}\right)=\alpha^{3}\beta\left(1+\frac{2\beta}{\alpha}r+\frac{\beta^{2}}{\alpha^{2}}r^{2}\right)p.
\end{equation}
We see that $r$ picks up a time dependence due to the pressure. This could have interesting phenomenological consequences, which we will not discuss here. An exception is the special parameter choice
\begin{equation}
\beta_{2}=\beta_{1}\frac{\beta}{\alpha},\qquad\beta_{3}=\beta_{1}\frac{\beta^{2}}{\alpha^{2}},
\end{equation}
which includes the vacuum energy case discussed in \cref{sec:vac}. For these parameters, the constraint becomes
\begin{equation}
\frac{p}{M_\eff^{2}}=\frac{m^{2}\beta_{1}}{\alpha^{3}\beta},
\end{equation}
i.e., $p$ is fixed to a constant, parameter-dependent value. This would correspond to a vacuum-type energy, $w=-1$, with a specific density and pressure. If the Universe contains any matter beyond that, then this algebraic branch is inconsistent and solutions will pick out the dynamical branch of the Bianchi constraint instead.

A special case when $p\neq0$ is if the equation of state is exactly $w=-1/3$, i.e., pure curvature, for which the right hand side of \cref{eq:firstbrancheq} becomes constant. A nontrivial solution to this for $\beta_1=\beta_2=0$ is to have a constant $a_f$, i.e., massive gravity in which the $f$-metric is fixed, as discussed in \rcite{Solomon:2014iwa,Gumrukcuoglu:2014xba}.

\bibliographystyle{JHEP}
\bibliography{bibliography}{}

\providecommand{\href}[2]{#2}\begingroup\raggedright\begin{thebibliography}{10}

\bibitem{deRham:2010ik}
C.~de~Rham and G.~Gabadadze, {\it {Generalization of the Fierz-Pauli Action}},
  {\em Phys.Rev.} {\bf D82} (2010) 044020,
  [\href{http://xxx.lanl.gov/abs/1007.0443}{{\tt arXiv:1007.0443}}].

\bibitem{deRham:2010kj}
C.~de~Rham, G.~Gabadadze, and A.~J. Tolley, {\it {Resummation of Massive
  Gravity}},  {\em Phys.Rev.Lett.} {\bf 106} (2011) 231101,
  [\href{http://xxx.lanl.gov/abs/1011.1232}{{\tt arXiv:1011.1232}}].

\bibitem{deRham:2011rn}
C.~de~Rham, G.~Gabadadze, and A.~J. Tolley, {\it {Ghost free Massive Gravity in
  the St\"uckelberg language}},  {\em Phys.Lett.} {\bf B711} (2012) 190--195,
  [\href{http://xxx.lanl.gov/abs/1107.3820}{{\tt arXiv:1107.3820}}].

\bibitem{deRham:2011qq}
C.~de~Rham, G.~Gabadadze, and A.~J. Tolley, {\it {Helicity Decomposition of
  Ghost-free Massive Gravity}},  {\em JHEP} {\bf 1111} (2011) 093,
  [\href{http://xxx.lanl.gov/abs/1108.4521}{{\tt arXiv:1108.4521}}].

\bibitem{Hassan:2011vm}
S.~Hassan and R.~A. Rosen, {\it {On Non-Linear Actions for Massive Gravity}},
  {\em JHEP} {\bf 1107} (2011) 009,
  [\href{http://xxx.lanl.gov/abs/1103.6055}{{\tt arXiv:1103.6055}}].

\bibitem{Hassan:2011hr}
S.~Hassan and R.~A. Rosen, {\it {Resolving the Ghost Problem in non-Linear
  Massive Gravity}},  {\em Phys.Rev.Lett.} {\bf 108} (2012) 041101,
  [\href{http://xxx.lanl.gov/abs/1106.3344}{{\tt arXiv:1106.3344}}].

\bibitem{Hassan:2011tf}
S.~Hassan, R.~A. Rosen, and A.~Schmidt-May, {\it {Ghost-free Massive Gravity
  with a General Reference Metric}},  {\em JHEP} {\bf 1202} (2012) 026,
  [\href{http://xxx.lanl.gov/abs/1109.3230}{{\tt arXiv:1109.3230}}].

\bibitem{Hassan:2011zd}
S.~Hassan and R.~A. Rosen, {\it {Bimetric Gravity from Ghost-free Massive
  Gravity}},  {\em JHEP} {\bf 1202} (2012) 126,
  [\href{http://xxx.lanl.gov/abs/1109.3515}{{\tt arXiv:1109.3515}}].

\bibitem{deRham:2014zqa}
C.~de~Rham, {\it {Massive Gravity}},
  \href{http://xxx.lanl.gov/abs/1401.4173}{{\tt arXiv:1401.4173}}.

\bibitem{Boulware:1973my}
D.~Boulware and S.~Deser, {\it {Can gravitation have a finite range?}},  {\em
  Phys.Rev.} {\bf D6} (1972) 3368--3382.

\bibitem{Akrami:2013ffa}
Y.~Akrami, T.~S. Koivisto, D.~F. Mota, and M.~Sandstad, {\it {Bimetric gravity
  doubly coupled to matter: theory and cosmological implications}},  {\em JCAP}
  {\bf 1310} (2013) 046, [\href{http://xxx.lanl.gov/abs/1306.0004}{{\tt
  arXiv:1306.0004}}].

\bibitem{Akrami:2014lja}
Y.~Akrami, T.~S. Koivisto, and A.~R. Solomon, {\it {The nature of spacetime in
  bigravity: two metrics or none?}},
  \href{http://xxx.lanl.gov/abs/1404.0006}{{\tt arXiv:1404.0006}}.

\bibitem{Hassan:2012wr}
S.~Hassan, A.~Schmidt-May, and M.~von Strauss, {\it {On Consistent Theories of
  Massive Spin-2 Fields Coupled to Gravity}},
  \href{http://xxx.lanl.gov/abs/1208.1515}{{\tt arXiv:1208.1515}}.

\bibitem{Yamashita:2014fga}
Y.~Yamashita, A.~De~Felice, and T.~Tanaka, {\it {Appearance of Boulware-Deser
  ghost in bigravity with doubly coupled matter}},
  \href{http://xxx.lanl.gov/abs/1408.0487}{{\tt arXiv:1408.0487}}.

\bibitem{deRham:2014naa}
C.~de~Rham, L.~Heisenberg, and R.~H. Ribeiro, {\it {On couplings to matter in
  massive (bi-)gravity}},  \href{http://xxx.lanl.gov/abs/1408.1678}{{\tt
  arXiv:1408.1678}}.

\bibitem{deRham:2014fha}
C.~de~Rham, L.~Heisenberg, and R.~H. Ribeiro, {\it {Ghosts \& Matter Couplings
  in Massive (bi-\&multi-)Gravity}},
  \href{http://xxx.lanl.gov/abs/1409.3834}{{\tt arXiv:1409.3834}}.

\bibitem{Hassan:2014gta}
S.~Hassan, M.~Kocic, and A.~Schmidt-May, {\it {Absence of ghost in a new
  bimetric-matter coupling}},  \href{http://xxx.lanl.gov/abs/1409.1909}{{\tt
  arXiv:1409.1909}}.

\bibitem{Noller:2014ioa}
J.~Noller, {\it {On Consistent Kinetic and Derivative Interactions for
  Gravitons}},  \href{http://xxx.lanl.gov/abs/1409.7692}{{\tt
  arXiv:1409.7692}}.

\bibitem{Soloviev:2014eea}
V.~O. Soloviev, {\it {Bigravity in tetrad Hamiltonian formalism and matter
  couplings}},  \href{http://xxx.lanl.gov/abs/1410.0048}{{\tt
  arXiv:1410.0048}}.

\bibitem{Noller:2014sta}
J.~Noller and S.~Melville, {\it {The coupling to matter in Massive, Bi- and
  Multi-Gravity}},  \href{http://xxx.lanl.gov/abs/1408.5131}{{\tt
  arXiv:1408.5131}}.

\bibitem{Solomon:2014iwa}
A.~R. Solomon, J.~Enander, Y.~Akrami, T.~S. Koivisto, F.~K{\"o}nnig, and
  E.~M{\"o}rtsell, {\it {Does massive gravity have viable cosmologies?}},
  \href{http://xxx.lanl.gov/abs/1409.8300}{{\tt arXiv:1409.8300}}.

\bibitem{Gumrukcuoglu:2014xba}
A.~E. G{\"u}mr{\"u}k{\c c}{\"u}o{\u g}lu, L.~Heisenberg, and S.~Mukohyama, {\it
  {Cosmological perturbations in massive gravity with doubly coupled matter}},
  \href{http://xxx.lanl.gov/abs/1409.7260}{{\tt arXiv:1409.7260}}.

\bibitem{Volkov:2011an}
M.~S. Volkov, {\it {Cosmological solutions with massive gravitons in the
  bigravity theory}},  {\em JHEP} {\bf 1201} (2012) 035,
  [\href{http://xxx.lanl.gov/abs/1110.6153}{{\tt arXiv:1110.6153}}].

\bibitem{Comelli:2011zm}
D.~Comelli, M.~Crisostomi, F.~Nesti, and L.~Pilo, {\it {FRW Cosmology in Ghost
  Free Massive Gravity}},  {\em JHEP} {\bf 1203} (2012) 067,
  [\href{http://xxx.lanl.gov/abs/1111.1983}{{\tt arXiv:1111.1983}}].

\bibitem{vonStrauss:2011mq}
M.~von Strauss, A.~Schmidt-May, J.~Enander, E.~M{\"{o}}rtsell, and S.~Hassan,
  {\it {Cosmological Solutions in Bimetric Gravity and their Observational
  Tests}},  {\em JCAP} {\bf 1203} (2012) 042,
  [\href{http://xxx.lanl.gov/abs/1111.1655}{{\tt arXiv:1111.1655}}].

\bibitem{Berg:2012kn}
M.~Berg, I.~Buchberger, J.~Enander, E.~M{\"{o}}rtsell, and S.~Sjors, {\it
  {Growth Histories in Bimetric Massive Gravity}},  {\em JCAP} {\bf 1212}
  (2012) 021, [\href{http://xxx.lanl.gov/abs/1206.3496}{{\tt
  arXiv:1206.3496}}].

\bibitem{Akrami:2012vf}
Y.~Akrami, T.~S. Koivisto, and M.~Sandstad, {\it {Accelerated expansion from
  ghost-free bigravity: a statistical analysis with improved generality}},
  {\em JHEP} {\bf 1303} (2013) 099,
  [\href{http://xxx.lanl.gov/abs/1209.0457}{{\tt arXiv:1209.0457}}].

\bibitem{Akrami:2013pna}
Y.~Akrami, T.~S. Koivisto, and M.~Sandstad, {\it {Cosmological constraints on
  ghost-free bigravity: background dynamics and late-time acceleration}},
  \href{http://xxx.lanl.gov/abs/1302.5268}{{\tt arXiv:1302.5268}}.

\bibitem{Konnig:2013gxa}
F.~K{\"o}nnig, A.~Patil, and L.~Amendola, {\it {Viable cosmological solutions
  in massive bimetric gravity}},  {\em JCAP} {\bf 1403} (2014) 029,
  [\href{http://xxx.lanl.gov/abs/1312.3208}{{\tt arXiv:1312.3208}}].

\bibitem{Enander:2013kza}
J.~Enander and E.~M{\"{o}}rtsell, {\it {Strong lensing constraints on bimetric
  massive gravity}},  {\em JHEP} {\bf 1310} (2013) 031,
  [\href{http://xxx.lanl.gov/abs/1306.1086}{{\tt arXiv:1306.1086}}].

\bibitem{Solomon:2014dua}
A.~R. Solomon, Y.~Akrami, and T.~S. Koivisto, {\it {Linear growth of structure
  in massive bigravity}},  \href{http://xxx.lanl.gov/abs/1404.4061}{{\tt
  arXiv:1404.4061}}.

\bibitem{Comelli:2012db}
D.~Comelli, M.~Crisostomi, and L.~Pilo, {\it {Perturbations in Massive Gravity
  Cosmology}},  {\em JHEP} {\bf 1206} (2012) 085,
  [\href{http://xxx.lanl.gov/abs/1202.1986}{{\tt arXiv:1202.1986}}].

\bibitem{Konnig:2014xva}
F.~K{\"o}nnig, Y.~Akrami, L.~Amendola, M.~Motta, and A.~R. Solomon, {\it
  {Stable and unstable cosmological models in bimetric massive gravity}},
  \href{http://xxx.lanl.gov/abs/1407.4331}{{\tt arXiv:1407.4331}}.

\bibitem{Comelli:2014bqa}
D.~Comelli, M.~Crisostomi, and L.~Pilo, {\it {FRW Cosmological Perturbations in
  Massive Bigravity}},  \href{http://xxx.lanl.gov/abs/1403.5679}{{\tt
  arXiv:1403.5679}}.

\bibitem{DeFelice:2014nja}
A.~De~Felice, A.~E. Gumrukcuoglu, S.~Mukohyama, N.~Tanahashi, and T.~Tanaka,
  {\it {Viable cosmology in bimetric theory}},
  \href{http://xxx.lanl.gov/abs/1404.0008}{{\tt arXiv:1404.0008}}.

\bibitem{Konnig:2014dna}
F.~K{\"o}nnig and L.~Amendola, {\it {Instability in a minimal bimetric gravity
  model}},  {\em Phys.Rev.} {\bf D90} (2014) 044030,
  [\href{http://xxx.lanl.gov/abs/1402.1988}{{\tt arXiv:1402.1988}}].

\bibitem{Lagos:2014lca}
M.~Lagos and P.~G. Ferreira, {\it {Cosmological perturbations in massive
  bigravity}},  \href{http://xxx.lanl.gov/abs/1410.0207}{{\tt
  arXiv:1410.0207}}.

\bibitem{Tamanini:2013xia}
N.~Tamanini, E.~N. Saridakis, and T.~S. Koivisto, {\it {The Cosmology of
  Interacting Spin-2 Fields}},  \href{http://xxx.lanl.gov/abs/1307.5984}{{\tt
  arXiv:1307.5984}}.

\bibitem{Aoki:2014cla}
K.~Aoki and K.-i. Maeda, {\it {Dark matter in ghost-free bigravity theory: From
  a galaxy scale to the universe}},
  \href{http://xxx.lanl.gov/abs/1409.0202}{{\tt arXiv:1409.0202}}.

\bibitem{Heisenberg:2014rka}
L.~Heisenberg, {\it {Quantum corrections in massive bigravity and new effective
  composite metrics}},  \href{http://xxx.lanl.gov/abs/1410.4239}{{\tt
  arXiv:1410.4239}}.

\bibitem{Afshar:2014dta}
H.~R. Afshar, E.~A. Bergshoeff, and W.~Merbis, {\it {Interacting spin-2 fields
  in three dimensions}},  \href{http://xxx.lanl.gov/abs/1410.6164}{{\tt
  arXiv:1410.6164}}.

\bibitem{Schmidt-May:2014xla}
A.~Schmidt-May, {\it {Mass eigenstates in bimetric theory with matter
  coupling}},  \href{http://xxx.lanl.gov/abs/1409.3146}{{\tt arXiv:1409.3146}}.

\bibitem{Suzuki:2011hu}
N.~Suzuki, D.~Rubin, C.~Lidman, G.~Aldering, R.~Amanullah, et~al., {\it {The
  Hubble Space Telescope Cluster Supernova Survey: V. Improving the Dark Energy
  Constraints Above $z>1$ and Building an Early-Type-Hosted Supernova Sample}},
   {\em Astrophys.J.} {\bf 746} (2012) 85,
  [\href{http://xxx.lanl.gov/abs/1105.3470}{{\tt arXiv:1105.3470}}].

\bibitem{Ade:2013zuv}
{\bf Planck} Collaboration, P.~Ade et~al., {\it {Planck 2013 results. XVI.
  Cosmological parameters}},  {\em Astron.Astrophys.} (2014)
  [\href{http://xxx.lanl.gov/abs/1303.5076}{{\tt arXiv:1303.5076}}].

\bibitem{Anderson:2012sa}
L.~Anderson, E.~Aubourg, S.~Bailey, D.~Bizyaev, M.~Blanton, et~al., {\it {The
  clustering of galaxies in the SDSS-III Baryon Oscillation Spectroscopic
  Survey: Baryon Acoustic Oscillations in the Data Release 9 Spectroscopic
  Galaxy Sample}},  {\em Mon.Not.Roy.Astron.Soc.} {\bf 427} (2013), no.~4
  3435--3467, [\href{http://xxx.lanl.gov/abs/1203.6594}{{\tt
  arXiv:1203.6594}}].

\bibitem{Beutler:2011hx}
F.~Beutler, C.~Blake, M.~Colless, D.~H. Jones, L.~Staveley-Smith, et~al., {\it
  {The 6dF Galaxy Survey: Baryon Acoustic Oscillations and the Local Hubble
  Constant}},  {\em Mon.Not.Roy.Astron.Soc.} {\bf 416} (2011) 3017--3032,
  [\href{http://xxx.lanl.gov/abs/1106.3366}{{\tt arXiv:1106.3366}}].

\bibitem{Blake:2011en}
C.~Blake, E.~Kazin, F.~Beutler, T.~Davis, D.~Parkinson, et~al., {\it {The
  WiggleZ Dark Energy Survey: mapping the distance-redshift relation with
  baryon acoustic oscillations}},  {\em Mon.Not.Roy.Astron.Soc.} {\bf 418}
  (2011) 1707--1724, [\href{http://xxx.lanl.gov/abs/1108.2635}{{\tt
  arXiv:1108.2635}}].

\bibitem{Sollerman:2009yu}
J.~Sollerman, E.~M{\"{o}}rtsell, T.~Davis, M.~Blomqvist, B.~Bassett, et~al.,
  {\it {First-Year Sloan Digital Sky Survey-II (SDSS-II) Supernova Results:
  Constraints on Non-Standard Cosmological Models}},  {\em Astrophys.J.} {\bf
  703} (2009) 1374--1385, [\href{http://xxx.lanl.gov/abs/0908.4276}{{\tt
  arXiv:0908.4276}}].

\bibitem{deRham:2012kf}
C.~de~Rham and S.~Renaux-Petel, {\it {Massive Gravity on de Sitter and Unique
  Candidate for Partially Massless Gravity}},  {\em JCAP} {\bf 1301} (2013)
  035, [\href{http://xxx.lanl.gov/abs/1206.3482}{{\tt arXiv:1206.3482}}].

\bibitem{Hassan:2012gz}
S.~Hassan, A.~Schmidt-May, and M.~von Strauss, {\it {On Partially Massless
  Bimetric Gravity}},  \href{http://xxx.lanl.gov/abs/1208.1797}{{\tt
  arXiv:1208.1797}}.

\bibitem{vanDam:1970vg}
H.~van Dam and M.~Veltman, {\it {Massive and massless Yang-Mills and
  gravitational fields}},  {\em Nucl.Phys.} {\bf B22} (1970) 397--411.

\bibitem{Zakharov:1970cc}
V.~Zakharov, {\it {Linearized gravitation theory and the graviton mass}},  {\em
  JETP Lett.} {\bf 12} (1970) 312.

\bibitem{Vainshtein:1972sx}
A.~Vainshtein, {\it {To the problem of nonvanishing gravitation mass}},  {\em
  Phys.Lett.} {\bf B39} (1972) 393--394.

\bibitem{Hassan:2013pca}
S.~Hassan, A.~Schmidt-May, and M.~von Strauss, {\it {Higher Derivative Gravity
  and Conformal Gravity From Bimetric and Partially Massless Bimetric Theory}},
   \href{http://xxx.lanl.gov/abs/1303.6940}{{\tt arXiv:1303.6940}}.

\bibitem{Hassan:2014vja}
S.~Hassan, A.~Schmidt-May, and M.~von Strauss, {\it {Particular Solutions in
  Bimetric Theory and Their Implications}},
  \href{http://xxx.lanl.gov/abs/1407.2772}{{\tt arXiv:1407.2772}}.

\bibitem{deRham:2013wv}
C.~de~Rham, K.~Hinterbichler, R.~A. Rosen, and A.~J. Tolley, {\it {Evidence for
  and obstructions to nonlinear partially massless gravity}},  {\em Phys.Rev.}
  {\bf D88} (2013), no.~2 024003,
  [\href{http://xxx.lanl.gov/abs/1302.0025}{{\tt arXiv:1302.0025}}].

\bibitem{deRham:2012ew}
C.~de~Rham, G.~Gabadadze, L.~Heisenberg, and D.~Pirtskhalava, {\it
  {Nonrenormalization and naturalness in a class of scalar-tensor theories}},
  {\em Phys.Rev.} {\bf D87} (2013), no.~8 085017,
  [\href{http://xxx.lanl.gov/abs/1212.4128}{{\tt arXiv:1212.4128}}].

\bibitem{deRham:2013qqa}
C.~de~Rham, L.~Heisenberg, and R.~H. Ribeiro, {\it {Quantum Corrections in
  Massive Gravity}},  {\em Phys.Rev.} {\bf D88} (2013) 084058,
  [\href{http://xxx.lanl.gov/abs/1307.7169}{{\tt arXiv:1307.7169}}].

\bibitem{Deffayet:2012zc}
C.~Deffayet, J.~Mourad, and G.~Zahariade, {\it {A note on 'symmetric' vielbeins
  in bimetric, massive, perturbative and non perturbative gravities}},  {\em
  JHEP} {\bf 1303} (2013) 086, [\href{http://xxx.lanl.gov/abs/1208.4493}{{\tt
  arXiv:1208.4493}}].

\bibitem{Gratia:2013gka}
P.~Gratia, W.~Hu, and M.~Wyman, {\it {Self-accelerating Massive Gravity: How
  Zweibeins Walk through Determinant Singularities}},  {\em Class.Quant.Grav.}
  {\bf 30} (2013) 184007, [\href{http://xxx.lanl.gov/abs/1305.2916}{{\tt
  arXiv:1305.2916}}].

\bibitem{Hoek:1982za}
J.~Hoek, {\it {On The Deser-van Nieuwenhuizen Algebraic Vierbein Gauge}},  {\em
  Lett.Math.Phys.} {\bf 6} (1982) 49--55.

\end{thebibliography}\endgroup

\end{document}